\documentstyle[epsf,preprint,aps]{revtex}
\begin{document}

\title{Retarded thermal Greens functions and forward scattering
amplitudes at two loops} 

\vspace{1in}

\author{F. T. Brandt$^\dagger$, Ashok Das$^\ddagger$ and J. Frenkel$^\dagger$}
\address{$^\dagger$Instituto de F\'\i sica,
Universidade de S\~ao Paulo\\
S\~ao Paulo, SP 05315-970, BRAZIL\\
$^\ddagger$Department of Physics and Astronomy,
University of Rochester\\
Rochester, NY 14627-0171, USA}

\date{\today}

\maketitle 

\vspace{1in}

\begin{abstract}

In this paper, we extend our earlier one loop
analysis to two loops and give a simple diagrammatic
description for the 
retarded Greens functions at finite temperature, 
in terms of forward scattering 
amplitudes of on-shell thermal particles. 
We present a simple discussion, which can be easily generalized to 
any field theory, of the temperature dependent
parts of the retarded two and three point functions 
in scalar field theory and QED.
As an application of our result at two loops, we
show how the infrared singularities in the thermal part of the 
retarded photon self-energy, cancel in QED$_{2}$ in the limit of
vanishing electron mass.  
\end{abstract}



\section{Introduction} \label{intro}

\par

In recent years, there has been an increased interest in the study of field 
theories at finite temperature for a variety of reasons
\cite{gross:1981br,weldon:1982aqweldon:1983jn,kajantie:1985xx,braaten:1990mzbraaten:1990az,smilga:1996cm}.
This has led to a 
better understanding of the formalism of such theories at finite temperature. 
In particular, we understand now that, in addition to the conventional 
imaginary time formalism, physical phenomena at finite temperature can equally
well be described by the real time formalism 
where the time variable is not traded in for the temperature of 
the system \cite{kapusta:book89,lebellac:book96,das:book97}.

The brute force calculations in field theories at finite temperature are quite
tedious, particularly, 
in the real time formalism where there is a necessity to double the degrees of
freedom of a theory which leads to $(2^{n}-1)$ independent causal amplitudes 
at $n$-th order. Therefore, it is quite important to find simple 
calculational methods in such theories for them to be useful. Furthermore, 
even though there exist various formal arguments showing the equivalence of 
the two distinct formalisms \cite{landsman:1987uw,evans:1992kyevans:1993ak,baier:1994yh,carrington:1996rx}
(imaginary time and real time), it is still
necessary to check explicitly that the two formalisms give the same physical 
quantities, at least in lower orders.

Amplitudes calculated in the imaginary time formalism naturally give rise to 
retarded (advanced) amplitudes with appropriate analytic continuation to the 
Minkowski space. In an earlier paper \cite{brandt:1998gb},
we gave a very simple diagrammatic 
representation of such amplitudes (retarded) 
in the real time formalism, at one 
loop, which is calculationally quite trivial and which also explicitly 
verifies the agreement of the two formalisms to this order. In this paper, we 
would like to extend our earlier results to two loops. Namely, we would 
give a simple diagrammatic description of the retarded two loop amplitudes at
finite temperature in the real time formalism which is calculationally much 
simpler (than a brute force calculation) and which also explicitly verifies 
the agreement of the imaginary time and the real time formalisms up to two 
loops. This description expresses the temperature dependent part of 
the retarded Green functions in terms of forward scattering amplitudes
for on-shell particles of thermal medium. These amplitudes are
multiplied by the appropriate statistical factors and
integrated over the three-momenta of the thermal particles.

The paper is organized as follows. In section \ref{sec2}, we analyze the 
temperature dependent part of the  retarded self-energy for the $\phi^{4}$ 
theory up to two loops in the real time formalism and give a simple 
diagrammatic representation for it. We explicitly evaluate the retarded 
self-energy and show that it coincides with the results obtained earlier using
the imaginary time formalism. In section \ref{sec3}, we analyze the temperature
dependent part of the retarded photon self-energy in QED at two loops
and obtain a simple diagrammatic representation for it. We also analyze
briefly and find a simple diagrammatic representation for the
temperature dependent retarded three point function at two loops
in QED. (The  diagrams for the amplitudes, in this theory, are  more like those
in the $\phi^3$ model.) Since our discussion is quite general,
it is easy to extend these results to the temperature 
dependent part of any retarded $n$-point amplitude at two loops
(for any bosonic or fermionic theory). In section \ref{sec4}, 
we discuss, as an example of our method, how the infrared mass 
singularities in the retarded photon self-energy, 
cancel at finite temperature for the Schwinger model \cite{schwinger:1962tp}
at two loops.
Finally, we present a brief conclusion in section \ref{sec5}.

\section{Retarded Self-Energy in $\phi^{4}$ Theory}\label{sec2}

Let us consider the theory of a self-interacting real scalar field described 
by the Lagrangian density

\begin{equation}
{\cal L} = \frac{1}{2} \partial_{\mu}\phi\partial^{\mu}\phi - 
\frac{m^{2}}{2}\phi^{2} - \frac{\lambda}{4!}\phi^{4}\label{1}
\end{equation}
As is well known, at  finite temperature, in the real time formalism (we will
follow the closed time path formalism although everything can be carried over 
to thermo field dynamics equally well), one needs to double the degrees of 
freedom, which leads to a $2\times 2$ matrix structure for the propagator

\begin{equation}
G_{ab}(q) = G_{ab}^{(0)}(q) + G_{ab}^{(\beta)}(q)\hspace{.5in} a,b = \pm
\label{2}
\end{equation}
Here $G_{ab}^{(0)}$ represents the zero temperature propagator while 
$G_{ab}^{(\beta)}$ is the on-shell, temperature dependent part of the 
propagator and, in the closed time path formalism, all
the four components of the temperature dependent part of the propagator are 
identical. Explicitly,

\begin{eqnarray}
G_{}^{(0)}(q) & = & \left( \begin{array}{cc}
\frac{1}{q^{2}-m^{2}+i\epsilon} & -2i\pi\theta(-q^{0})\delta(
q^{2}-m^{2})\\
-2i\pi\theta(q^{0})\delta(q^{2}-m^{2}) & -\frac{1}{q^{2}-m^{2}-i\epsilon}
\end{array}\right)\nonumber\\
G_{}^{(\beta)}(q) & = & -2i\pi n_{B}(|q^{0}|)\delta(q^{2}-m^{2})
\left(\begin{array}{cc}
1 & 1\\
1 & 1
\end{array}\right)\label{3}
\end{eqnarray}
where $n_{B}(|q^{0}|)$ stands for the bosonic distribution function (and the
propagator really is $iG_{ab}$).
Following ref. \cite{brandt:1998gb},
we introduce the following diagrammatic representation for
the two parts of the propagator
\begin{equation}
G_{a b}^{(0)}(q)=
  \begin{array}[h]{c}
\epsfbox{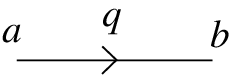}
\end{array};
\;\;\;\;\;\;\;
G_{a b}^{(\beta)}(q)=
  \begin{array}[h]{c}
\epsfbox{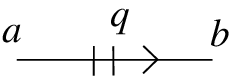}
\end{array} 
\end{equation}
and the temperature dependent part of the self-energy, at one loop,
is easily obtained to be 
(this also coincides with the retarded self-energy to this order)
the forward scattering amplitude for a single on-shell 
thermal particle 
\begin{equation}
  \begin{array}[h]{c}
\epsfbox{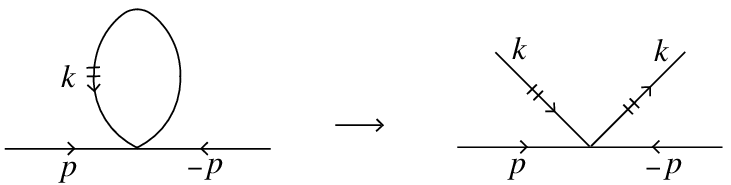}
  \end{array}
\end{equation}

In going beyond one loop, we note that 
there are two kinds of diagrams for the
self-energy at two loops in this theory, namely, the penguin and the 
rising sun diagrams. Let us analyze each class of the diagrams (which 
involve three internal propagators) separately.
For the penguins, it is straightforward to check that all the diagrams for the
retarded self-energy where all three internal propagators are cut cancel, 
namely,

\begin{equation}
  \begin{array}[h]{c}
    \epsfbox{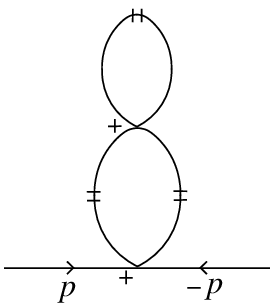}
 \end{array}\;\;\; +\;\;\;
  \begin{array}[h]{c}
    \epsfbox{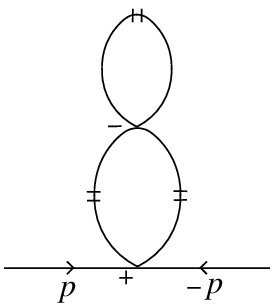}
 \end{array}\;\;\; =\;\;\; 0 .
\end{equation}
This follows trivially from the fact that the cut propagators are 
the same, independent of their thermal indices, as is clear from Eq. (\ref{3})
and the fact that the vertices with the $-$ thermal index have a  negative 
sign relative to those with the $+$ thermal index. Some of the diagrams with 
two internal cut propagators can also be easily seen to add up to zero
\begin{equation}
  \begin{array}[h]{c}
    \epsfbox{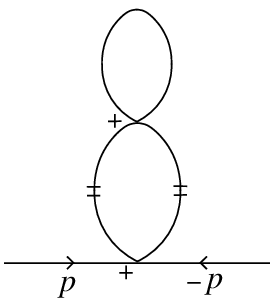}
   \end{array}\;\;\; +\;\;\;
  \begin{array}[h]{c}
    \epsfbox{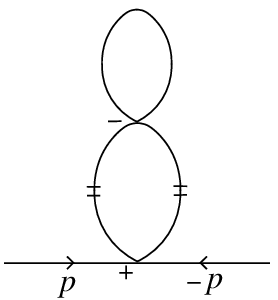}
   \end{array}\;\;\; =\;\;\; 0 .
\end{equation}

However, not all the diagrams with two internal cut propagators 
add up to zero and all the diagrams with a single cut propagator do 
contribute. Thus, all the temperature dependent contribution
coming from the penguin diagrams to 
the two loop retarded self-energy can be seen to be
\begin{eqnarray}
  \begin{array}[h]{c}
    \epsfbox{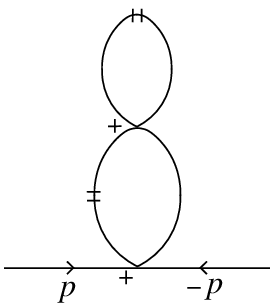}
  \end{array}\;\;\; +\;\;\;
  \begin{array}[h]{c}
    \epsfbox{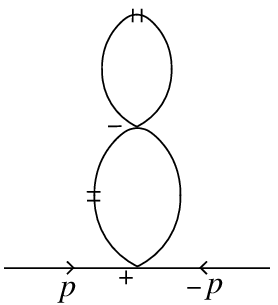}
  \end{array}\;\;\; +\;\;\; 
  \begin{array}[h]{c}
    \epsfbox{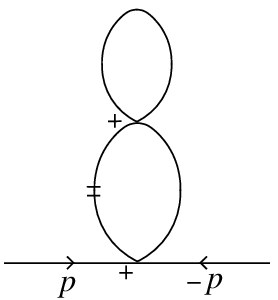}
  \end{array}\;\;\; +\;\;\; 
\nonumber \\
  \begin{array}[h]{c}
    \epsfbox{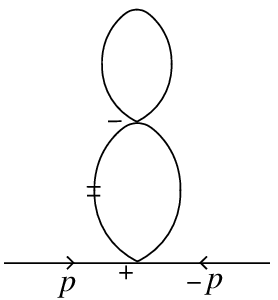}
  \end{array}\;\;\; +\;\;\;
  \begin{array}[h]{c}
    \epsfbox{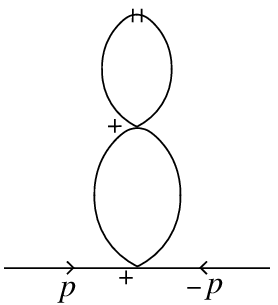}
  \end{array}\;\;\; +\;\;\; 
  \begin{array}[h]{c}
    \epsfbox{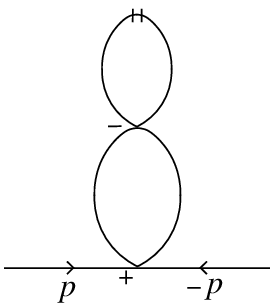}
  \end{array}\;\;\; \;\; \; \;\;\; ,
\end{eqnarray}
where graphs containing permuted cuts on the internal 
propagators in the lower bubble are to be understood.

Next, let us look at the rising sun diagrams for the self-energy at two 
loops.Once again, in the retarded self-energy, 
all the diagrams with three internal cut propagators add up to zero
\begin{equation}
  \begin{array}[h]{c}
    \epsfbox{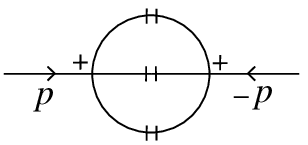}
   \end{array}\;\;\; +\;\;\;
  \begin{array}[h]{c}
    \epsfbox{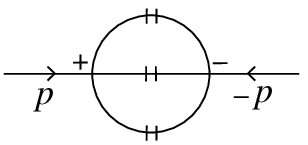}
   \end{array}\;\;\; =\;\;\; 0 
\end{equation}
which, again, follows trivially from the fact that all the cut 
propagators are the same independent of their thermal indices and that the
vertices with $-$ index have a relative negative sign. The diagrams with two
internal cut propagators as well as a single internal cut propagator, however,
all contribute, so that
\begin{eqnarray}
  \begin{array}[h]{c}
    \epsfbox{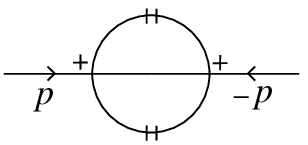}
  \end{array}\;\;\; +\;\;\;
  \begin{array}[h]{c}
    \epsfbox{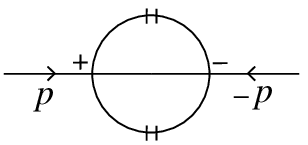}
  \end{array}\;\;\; +\;\;\; 
\nonumber \\
  \begin{array}[h]{c}
    \epsfbox{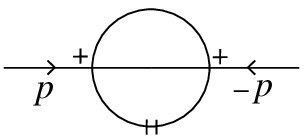}
  \end{array}\;\;\; +\;\;\;
  \begin{array}[h]{c}
    \epsfbox{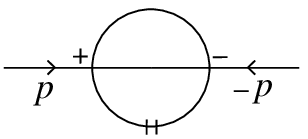}
  \end{array}\;\;\; \;\;\;\; \; \;\;\; 
\end{eqnarray}
is nonzero (diagrams with permuted cuts must also be considered).

We next recall that a cut propagator is an on-shell propagator and hence can
be thought of as a cut open external line representing an on-shell thermal 
particle (intuitively, one can think of it as a real particle of the medium). 
Therefore, we can combine the non-vanishing penguin and the rising 
sun diagrams to represent the temperature dependent part of the retarded
self-energy at two loops as
\begin{eqnarray}
\Sigma_{R}^{(\beta)}=
  \begin{array}[h]{c}
    \epsfbox{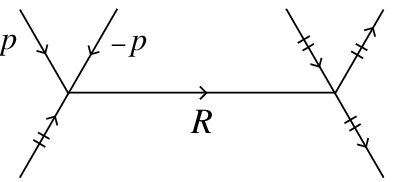}
  \end{array}\;\; + \;\; &
  \begin{array}[h]{c}
    \epsfbox{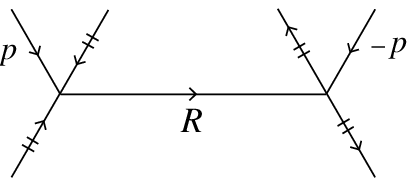}
  \end{array}\;\; +\;\; &
\nonumber \\
  \begin{array}[h]{c}
    \epsfbox{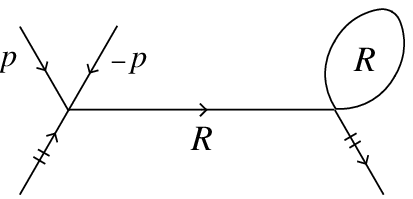}
  \end{array}\;\; +\;\; &
  \begin{array}[h]{c}
    \epsfbox{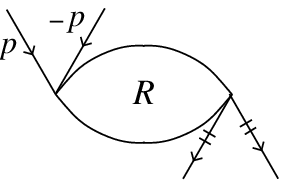}
  \end{array}\;\; +\;\; &
  \begin{array}[h]{c}
    \epsfbox{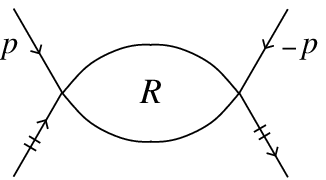}
  \end{array}
\end{eqnarray}
where we have used the convention that the cut lines on the 
lower/upper half come from the same propagator. Furthermore, $G_{R}$ 
represents the usual retarded propagator, namely,

\begin{equation}
G_{R}(q) = \frac{1}{q^{2}-m^{2}+i\epsilon q^{0}}
\end{equation}
and the definition of the retarded 
vertex functions are as given in ref. \cite{brandt:1998gb}.
Thus, we see that the temperature 
dependent part of the retarded self-energy at two loops has two kinds of terms
- terms quadratic in the statistical factor as well as ones linear in the 
statistical factor (remember that each cut propagator carries a statistical
factor). Furthermore, the terms quadratic in the statistical factor 
corresponding to two cut lines can be thought of  simply as the  forward 
scattering amplitude of two on-shell thermal particles.
The terms which are linear in the 
statistical factor or have a single cut propagator, on the other hand, 
correspond to the forward scattering amplitudes of a single on-shell thermal 
particle with all possible one loop self-energy and vertex 
corrections where the corrections are retarded and have to be evaluated
at zero temperature. 

All these terms are rather simple to evaluate, by doing the
energy integrals with the help of the delta functions in Eq. (\ref{3}).
One then obtains from the 
diagrams which depend on the external momentum,
that the relevant temperature dependent part at two loops is given by
\begin{eqnarray}
\Sigma_{R}^{(\beta)}(p_0,\vec p) &=&
\frac{\lambda^2}{16}\int
\frac{d^3 k}{\left(2\pi\right)^3}
\frac{d^3 q}{\left(2\pi\right)^3}
\frac{1}{\omega_k}\frac{1}{\omega_q}\frac{1}{\omega_r}
\left\{n_B(\omega_{k})\left[
D(\omega_{k},\omega_{q},\omega_{r})+
D(-\omega_{ k},\omega_{ q},\omega_{ r})\right]\right.
\nonumber \\
&+&n_B(\omega_{ k}) n_B(\omega_{ q})\left[
D(\omega_{ k},\omega_{ q},\omega_{ r})+
D(-\omega_{ k},\omega_{ q},\omega_{ r})\right.
\nonumber \\
&+&
\left.\left.
D(\omega_{ k},-\omega_{ q},\omega_{ r})-
D(\omega_{ k},\omega_{ q},-\omega_{ r})
\right]
\right\}
\label{4},
\end{eqnarray}
where $\omega_{ k}\equiv\sqrt{\vec k^2+m^2}$,  \,
$\vec r = \vec k + \vec q - \vec p$ and
\begin{equation}
D(\omega_{k},\omega_{ q},\omega_{ r})=
\frac{1}{p_0+i\epsilon + \omega_{ k} + \omega_{ q}
+ \omega_{ r}} +
\frac{1}{-(p_0+i\epsilon) + \omega_{ k} + \omega_{ q}
+ \omega_{ r}}.
\label{4a}
\end{equation}
Such a form of the retarded thermal self-energy is rather convenient,
particularly for the calculation of its imaginary part which is
related to the plasma damping rate.
The above result agrees with the one obtained by Parwani
in ref. \cite{parwani:1992gq}, 
after analytic continuation 
from the imaginary time formalism. 
In this way, one can establish an explicit equivalence between
the two formalisms up to two loops much like in the one loop case.

\section{Retarded Greens functions in QED}\label{sec3}

Let us next consider the QED theory described by
the Lagrangian density
\begin{equation}
{\cal L} = - \frac{1}{4} F_{\mu\nu}F^{\mu\nu} + \overline{\psi}
(i\partial\!\!\!\slash - m + eA\!\!\!\slash)
\psi,\label{5}
\end{equation}
where $A_{\mu}$ is the photon field, $F_{\mu\nu}$ the field strength 
tensor and $\psi$ is the Dirac field.
Apart from a matrix factor and the replacement of
$(-n_B)$ by the fermionic distribution function $n_F$,
the structure of the electron propagator is similar to that
in Eqs. (\ref{2}) and (\ref{3}). 
However, let us note that, for the 
purpose of our discussion, the actual form of the propagator is irrelevant, 
all that is crucial is that the temperature dependent part of any propagator 
(bosonic or fermionic) is the same for all the thermal indices in the closed 
time path formalism. (Consequently, our  discussion will be quite 
general without restrictions to a particular theory.) 
Assuming that the largest time vertex is the one conected with the
external line $p$, 
the retarded thermal self-energy 
of photons at one loop is given by the
forward scattering amplitude of a single on-shell thermal particle
\begin{equation}
\left[   \begin{array}[h]{c}
\epsfbox{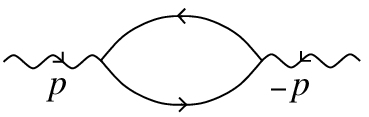} 
  \end{array}
\right]_{Ret}^{\beta}
\longrightarrow
\left.
  \begin{array}[h]{c}
\epsfbox{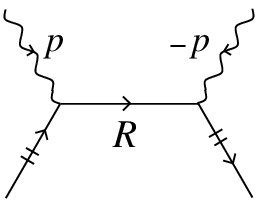} 
  \end{array}
+
  \begin{array}[h]{c}
\epsfbox{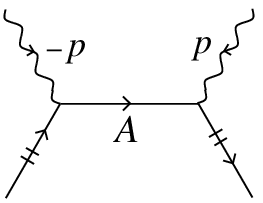} 
  \end{array}
\right.
\label{OneLoopQED}
\end{equation}
Note that the propagator with momentum flowing outwards from
the largest time vertex is retarded while the one with momentum
flowing towards this vertex is advanced
(see ref. \cite{brandt:1998gb} for details).

In going beyond one loop, we note that there are again two classes of diagrams
- ones containing overlapping divergences and the others with the structure 
of one loop diagrams with a self-energy correction for one of the internal 
propagators. Each of these classes of diagrams contains five 
internal propagators. As before, it is easy to check that the sum of all 
diagrams containing five cut propagators, four cut propagators as well as
three cut propagators individually vanish which follows from the fact that the
cut propagators are the same irrespective of their thermal indices and that 
vertices with $-$ thermal index have an extra negative sign. Some of the 
graphs with two internal cut propagators also cancel, but not all of them. And
all the graphs with a single cut propagator contribute. Adding them all up and
remembering that a cut propagator is on-shell and hence can be thought of as 
an external thermal particle, we again find that the temperature dependent 
part of the retarded self-energy, at two loops, can be written as
\begin{eqnarray}
&
\left[
  \begin{array}[h]{c}
\epsfbox{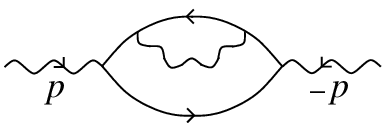} 
  \end{array}
+
  \begin{array}[h]{c}
\epsfbox{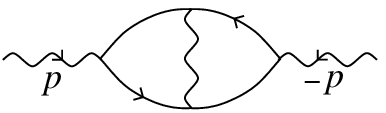} 
  \end{array}
\right]^{(\beta)}_{Ret}
\longrightarrow 
&
\nonumber \\
&
  \begin{array}[h]{c}
\\ \epsfbox{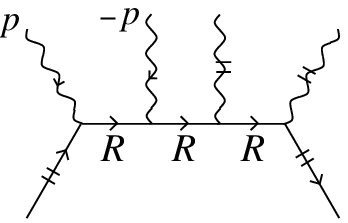}  \\ {\rm(a)}
  \end{array}
+
  \begin{array}[h]{c}
\\ \epsfbox{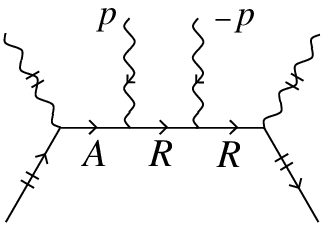}  \\ {\rm(b)}
  \end{array}
+
  \begin{array}[h]{c}
\\ \epsfbox{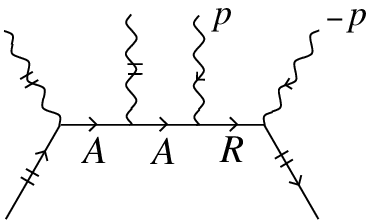}  \\ {\rm(c)}
  \end{array}
+
&
\nonumber \\
&
  \begin{array}[h]{c}
\\ \epsfbox{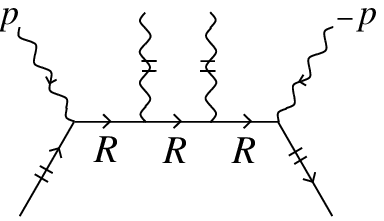} \\ {\rm(d)}
  \end{array}
+
  \begin{array}[h]{c}
\\ \epsfbox{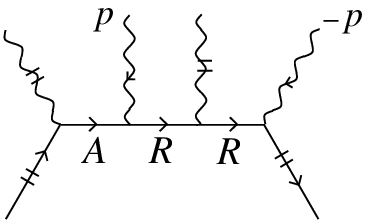} \\ {\rm(e)}
  \end{array}
+
  \begin{array}[h]{c}
\\ \epsfbox{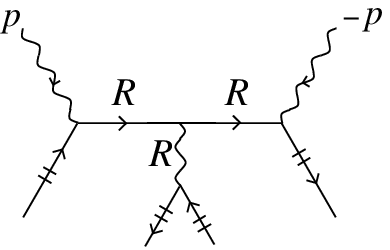} \\ {\rm(f)}
  \end{array}
+
&
\nonumber \\
&
\begin{array}[h]{c}
 \epsfbox{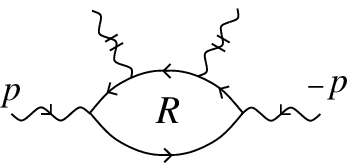} \\ {\rm(g)}
  \end{array}
+
  \begin{array}[h]{c}
\\
\epsfbox{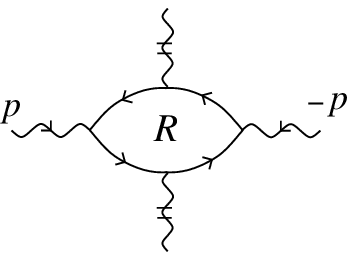} \\ {\rm(h)}
  \end{array}
+
  \begin{array}[h]{c}
\\ \epsfbox{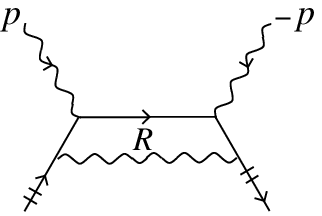} \\ {\rm(i)}
  \end{array}
+
&
\nonumber \\
&
  \begin{array}[h]{c}
\\ \epsfbox{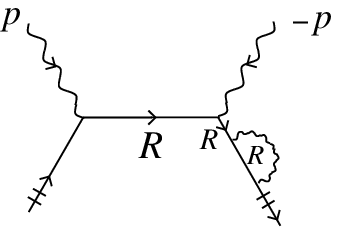} \\ {\rm(j)}
  \end{array}
+
  \begin{array}[h]{c}
\\ \epsfbox{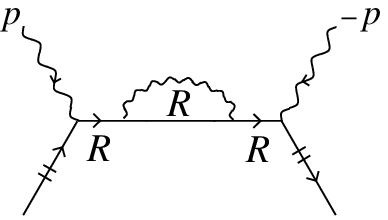} \\ {\rm(k)}
  \end{array}
+
  \begin{array}[h]{c}
\\ \epsfbox{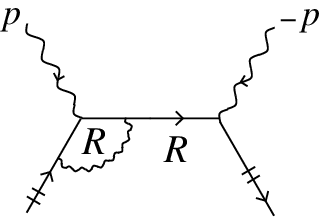} \\ {\rm(l)}
  \end{array}
&
\label{ForwardTwoLoopQED}
\end{eqnarray}
Here, all other diagrams obtained by the permutation of external lines
are to be understood.
Once again, it is clear that the temperature dependent part of the two loop 
retarded self-energy contains terms which are quadratic in the statistical 
factors as well as ones which are linear in the statistical factor. The terms
which are quadratic in the statistical factor, come from diagrams with two 
cut propagators each of which  can have a simple description of a  forward 
scattering amplitude for two on-shell thermal particles. 
On the other hand, those which are linear in the statistical 
factor, come from diagrams with a single cut propagator and can be described
in terms of  forward scattering amplitudes of a single on-shell thermal 
particle with all possible one loop self-energy and vertex
corrections. As in the last section, we note that these corrections are 
retarded and are to be evaluated at zero temperature. This is, indeed, a very
simple description of the two loop retarded self-energy and much simpler to 
evaluate than the brute force calculation would entail.

Let us evalulate next the temperature dependent part of the retarded three 
point function, at two loops order (although the three photon vertex
vanishes in QED by charge conjugation, our analysis does also
apply to
a theory like QCD, where the three gluon vertex is non-zero).
Without going into too much detail, let us note that there will be two classes 
of diagrams contributing to this - one containing diagrams which can be 
thought of as a vertex correction to the one loop three point function and 
the other with graphs which contain a one loop self-energy correction in one
of the internal propagators of the one loop three point vertex function. The 
diagrams in each
of these two classes of graphs consist of six internal propagators and it is 
straightforward to check that the sum of graphs with six cut propagators, five
cut propagators, four cut propagators as well as three cut propagators 
individually 
vanish. Some of the graphs with two cut propagators also add up to zero, but 
not all. Furthermore, all the graphs with a single cut propagator contribute. 
The non-vanishing diagrams 
can again be given a simple graphical representation 
by remembering that a cut line is on-shell and can be thought of as an 
external thermal particle (of the medium).
We thus get, for example, the graphical equations
\begin{eqnarray}
&
\left[
  \begin{array}[h]{c}
\epsfbox{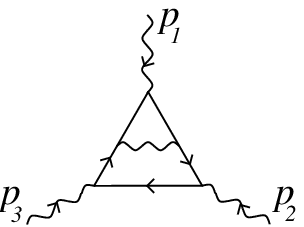}
  \end{array}
\right]^{(\beta)}_{Ret}
\longrightarrow 
  \begin{array}[h]{c}
\\ \epsfbox{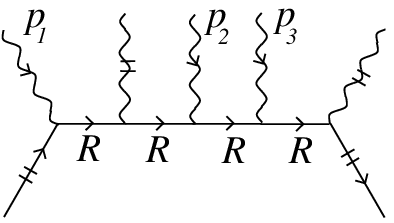}  \\ {\rm(a)}
  \end{array}
+
&
\nonumber \\
&
  \begin{array}[h]{c}
\\ \epsfbox{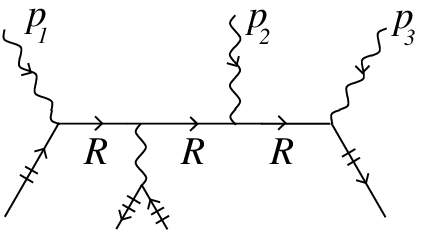}  \\ {\rm(b)}
  \end{array}
+
  \begin{array}[h]{c}
\\ \epsfbox{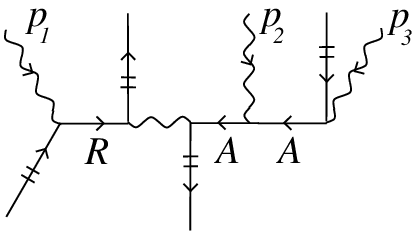}  \\ {\rm(c)}
  \end{array}
+
\nonumber \\
&
  \begin{array}[h]{c}
\\ \epsfbox{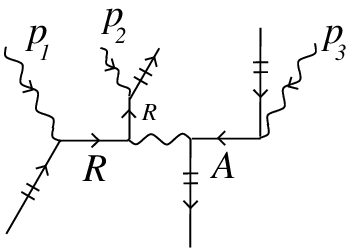}  \\ {\rm(d)}
  \end{array}
+
  \begin{array}[h]{c}
\\ \epsfbox{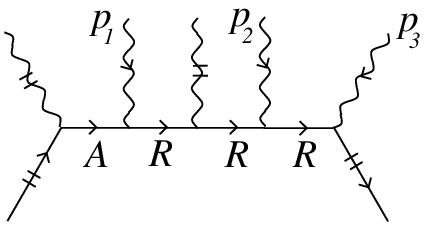}  \\ {\rm(e)}
  \end{array}
+
\nonumber \\
&
  \begin{array}[h]{c}

\\ \epsfbox{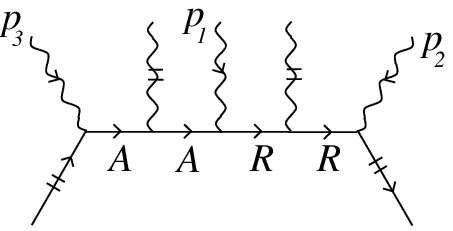}  \\ {\rm(f)}
  \end{array}
+
  \begin{array}[h]{c}
\\ \epsfbox{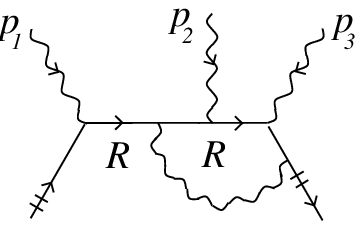}  \\ {\rm(g)}
  \end{array}
+
\nonumber \\
&
  \begin{array}[h]{c}
\\ \epsfbox{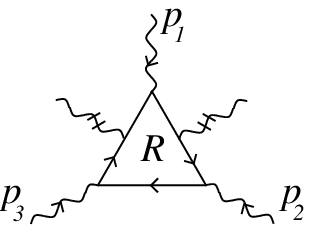}  \\ {\rm(h)}
  \end{array}
+
  \begin{array}[h]{c}
\\ \epsfbox{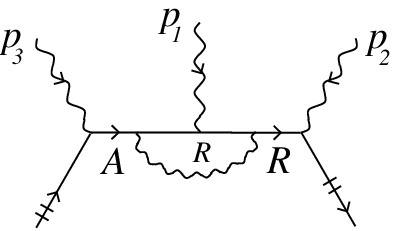}  \\ {\rm(i)}
  \end{array}
+
  \begin{array}[h]{c}
\\ \epsfbox{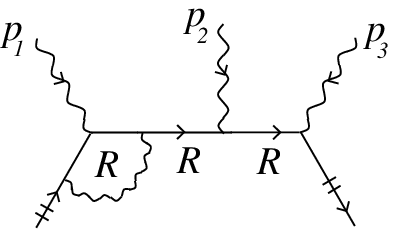}  \\ {\rm(j)}
  \end{array}
\end{eqnarray}

\begin{eqnarray}
&
\left[
  \begin{array}[h]{c}
\epsfbox{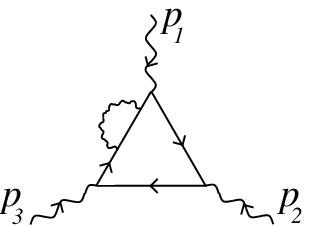}
  \end{array}
\right]^{(\beta)}_{Ret}
\longrightarrow 
  \begin{array}[h]{c}
\\ \epsfbox{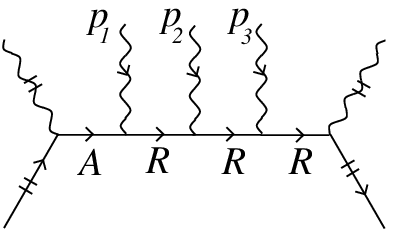}  \\ {\rm(a)}
  \end{array}
+
&
\nonumber \\
&
  \begin{array}[h]{c}
\\ \epsfbox{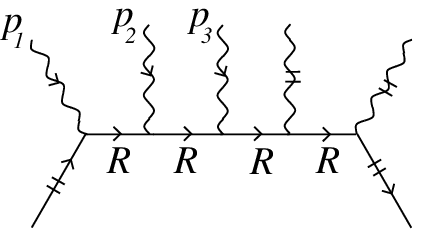}  \\ {\rm(b)}
  \end{array}
+
  \begin{array}[h]{c}
\\ \epsfbox{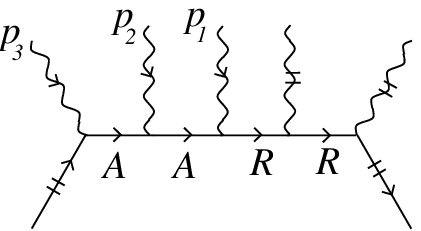}  \\ {\rm(c)}
  \end{array}
+
\nonumber \\
&
  \begin{array}[h]{c}
\\ \epsfbox{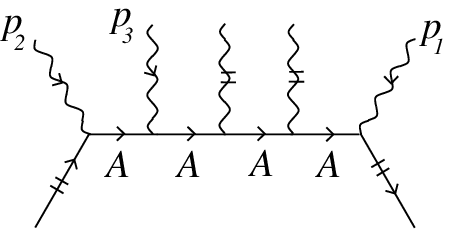}  \\ {\rm(d)}
  \end{array}
+
  \begin{array}[h]{c}
\\ \epsfbox{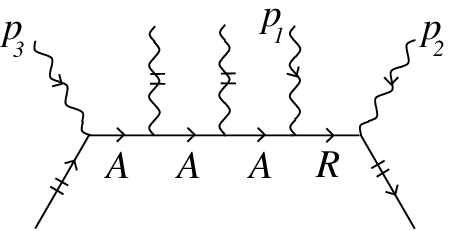}  \\ {\rm(e)}
  \end{array}
+
\nonumber \\
&
  \begin{array}[h]{c}

\\ \epsfbox{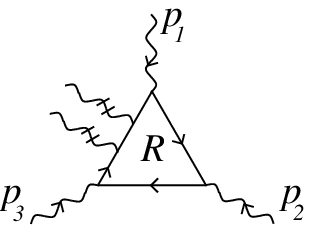}  \\ {\rm(f)}
  \end{array}
+
  \begin{array}[h]{c}
\\ \epsfbox{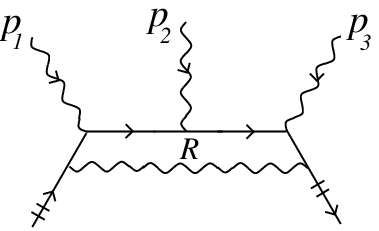}  \\ {\rm(g)}
  \end{array}
+
\nonumber \\
&
  \begin{array}[h]{c}
\\ \epsfbox{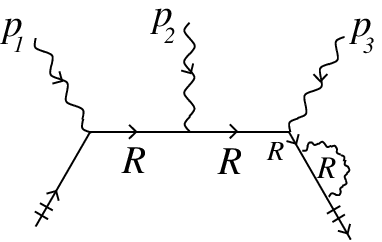}  \\ {\rm(h)}
  \end{array}
+
  \begin{array}[h]{c}
\\ \epsfbox{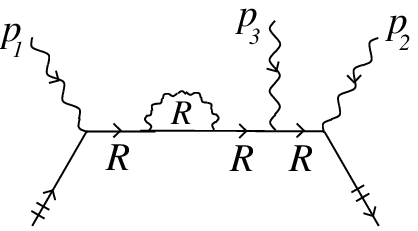}  \\ {\rm(i)}
  \end{array}
+
  \begin{array}[h]{c}
\\ \epsfbox{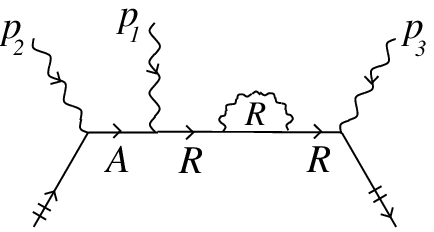}  \\ {\rm(j)}
  \end{array}
\end{eqnarray}
In the above forward scattering amplitudes, all other relevant
permutations of the external lines are to be understood.
Here, we have assumed that the vertex with external momentum $p_1$ 
corresponds to the largest time vertex.
Then, all propagators with momentum flowing towards this vertex
are advanced, whereas the propagators with momentum flowing outward
from the largest time vertex are retarded.
Once again, we note the familiar pattern. The temperature dependent
part of the retarded three point function, at two loops, consists of two types
of terms. The ones which are quadratic in the statistical factor correspond to
simple forward scattering amplitudes of two  on-shell thermal particles (of the
medium) whereas the ones which are linear in the statistical factor correspond
to the forward scattering amplitude of a single thermal on-shell particle
with all possible one loop self-energy and vertex 
corrections. Such retarded one loop corrections, 
which are calculated at zero temperature, may
contain ultraviolet divergences and must be renormalized as usual.

From the analysis of these two loop amplitudes so far, it is quite  
straightforward to find a 
simple, general diagrammatic description of the two loop retarded $n$-point 
amplitudes. {\it The temperature dependent part of a  general two loop 
retarded $n$-point amplitude can be written in terms of two classes of 
diagrams. The first class would involve simple forward scattering amplitudes 
for two on-shell thermal particles (two cut lines) and would correspond to 
terms which are quadratic in the statistical factor. The second class of 
diagrams would involve forward scattering amplitudes of a single on-shell 
thermal particle (one cut line) with all possible one loop self-energy
and vertex corrections, where these corrections are retarded 
and are to be evaluated at zero temperature.}

\section{Application}\label{sec4}

In this section, we will study the QED$_2$ model at finite 
temperature and, as an application of the general method, show how the 
infrared mass singularities
of the temperature dependent retarded photon self-energy 
cancel at two loops in the limit of vanishing fermion mass. This  
gauge theory, which is described by the Lagrangian density 
(\ref{5}) in $1+1$ dimensions, is well behaved in the ultraviolet region.
For simplicity, we will work in the Feynman gauge although 
the choice of gauge is immaterial, 
since the finite temperature results are gauge invariant
in the Schwinger model.

If we look at the photon self-energy in this theory, then, from our 
discussions, we recognize that the temperature dependent part of the retarded
self-energy, at one loop, can be written in terms of the forward scattering
amplitude of a thermal on-shell fermion as shown in
Eq. (\ref{OneLoopQED}).
The two diagrams give identical contribution, since inverting
the direction of the internal momentum flow interchanges the
advanced and retarded propagators. Unlike in 
higher dimensions, in $1+1$ dimensions, there is only one transverse structure 
for the polarization tensor at finite temperature. Furthermore, there are 
special identities for the Dirac trace in $1+1$ dimensions 
which make the one loop calculation quite simple and give
\begin{eqnarray}
\Pi_{R{(1)}}^{(\beta)}(p^{0},p^1) & = & 
-\frac{2e^{2}m^{2}}{\pi}\int d^{2}k\,\frac
{n_{F}(|k^{0}|)}{(k^1+p^1)^{2}-m^{2}+i\epsilon (k^{0}+p^{0})}\delta(k^{2}-m^{2})
\nonumber \\
 & = & -\frac{e^{2}m^{2}}{2\pi}\int dk^1\,\frac{n_{F}(\omega_{k})}{\omega_{k}
\omega_{k+p}}\left[\frac{1}{\omega_{k}+p^{0}-\omega_{k+p}+i\epsilon}-\frac
{1}{\omega_{k}+p^{0}+\omega_{k+p}+i\epsilon}\right.\nonumber\\
 &   &\hspace{.5in}\left. +\frac{1}{\omega_{k}-p^{0}-\omega_{k+p}-i\epsilon}-
\frac{1}{\omega_{k}-p^{0}+\omega_{k+p}-i\epsilon}\right]\label{7}
\end{eqnarray}
Here, we have identified the space components of $p^{\mu}$ and $k^{\mu}$ with 
$p^1$ and $k^1$ respectively, 
$n_{F}$ is the fermionic distribution function and
$\omega_{k} = \sqrt{(k^1)^{2}+m^{2}}$. 
Clearly, this vanishes as $m\rightarrow 0$.

In going beyond one loop, let us note that the infrared mass 
singularities in the 
retarded photon self-energy come from the graphs with a one loop 
self-energy correction in the internal fermion propagator, particularly when 
the internal photon propagator is cut (irrespective of whether there are other
cuts in the diagram).
Consider, for example, the following set of thermal diagrams
\begin{eqnarray}\label{Infrared}
\left[
  \begin{array}[h]{c}
\epsfbox{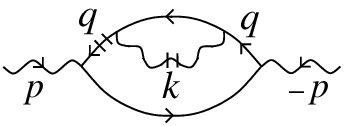}
  \end{array} \right. & 
+
  \begin{array}[h]{c}
\epsfbox{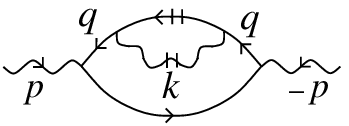}
  \end{array} &
+
\left.
  \begin{array}[h]{c}
\epsfbox{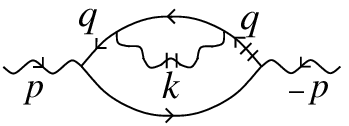}
  \end{array}
\right]_{Ret}
\\
{\rm(a)}\;\;\;\;\;\;\;\;\;\;\;\;\;\;\; &
\;\;\;\;\;{\rm(b)} &
\;\;\;\;\;\;\;\;\;\;\;\;\;\;\;\;\;{\rm(c)}
\nonumber
\end{eqnarray}
The cut photon line would give an infrared divergent
contribution of the form (coming from $k\rightarrow 0$)
\begin{equation}
I(T/\lambda)=
\int \frac{d^{2}k}{(2\pi)^{2}}\frac{1}{e^{|k_0|/T}-1}\delta(k^{2}-\lambda^{2})
 =\frac{1}{(2\pi)^{2}}
\left[\frac{\pi T}{2\lambda}+\frac{1}{2}\ln\frac{\lambda}{T}+\cdots\right]
\label{8}
\end{equation}
Here, $\lambda$ is an infrared regulating photon mass which has to be set to 
zero at the end. Once we have isolated this factor, the photon momentum can be 
safely set to zero in the rest of the diagrams.

As we have seen earlier, at 
two loops, only graphs with two cut propagators or a single cut propagator can
give a non-vanishing contribution. A little bit of analysis shows that in 
graphs with a single cut on the photon
propagator, which as shown in Fig. (\ref{ForwardTwoLoopQED}-g)
contain a retarded fermion loop
calculated at zero temperature, the infrared divergent contributions
cancel. The graphs with two cut 
propagators, shown in the set (\ref{Infrared}),
can be expressed in terms of the forward scattering 
amplitudes (a), (b) and (c), indicated in the graphical equation
(\ref{ForwardTwoLoopQED}).
To evaluate these, we note that in consequence of special kinematic
relations which hold in $1+1$ dimensions, there will again appear
in the numerator an overall factor $m^2$ (compare with Eq.
(\ref{7})). One may naively think that also in this case, this
factor would be sufficient to ensure the vanishing of the whole
contribution, in the limit $m\rightarrow 0$. However,such an argument
would be fallacious at two loop order, where the infrared
mass singularities
of the theory begin to manifest. In this case, there are individual
contributions whose denominators behave like $1/m^2$ as 
$m\rightarrow 0$.
Such terms would then give rise to contributions 
proportional to $I(T/\lambda)$, which are independent of $m$.
Therefore, a more complete analysis
is necessary in order to ascertain the behavior of the theory in
the limit of vanishing fermion mass.

To this end, we note that the infrared behavior of the diagrams
in the set (\ref{Infrared})
is determined by the regions $k\rightarrow 0$ and 
$q,\, m \ll p$ (we assume that $p^2\neq 0$). Then, in the
corresponding forward scattering amplitudes 
(\ref{ForwardTwoLoopQED}-a, b, c)), we expand the
retarded and advanced propagators into a sum of principal values
and delta functions. Adding all such contributions, it turns
out that the infrared divergent part of the above diagrams is 
given by
\begin{eqnarray}
\tilde\Pi_{R{(2)}}^{(\beta)}(\lambda,p) & = & 
\frac{16e^4}{p^2} m^2 I(T/\lambda)
\int d^{2}q\,
n_{F}(|q^{0}|) 
\left(q^{2}+m^2\right)\nonumber \\
& &
\left\{
\pi^2\left[\delta(q^2-m^2)\right]^3 -
3\left[{\cal P}\left(\frac{1}{q^2-m^2}\right)\right]^2
\delta(q^2-m^2)\right\}.
\label{9}
\end{eqnarray}
In order to simplify the expression in the curly bracket, we
make use of the identity
\begin{equation}
\frac{1}{(q^2-m^2+i\epsilon)^3}=
{\cal P}\left[\left(\frac{1}{q^2-m^2}\right)^3\right]
-\frac{i\pi}{2}\delta^{\prime\prime}(q^2-m^2),
\end{equation}
where the derivatives in the delta function are with
respect to its argument. From the imaginary part of this
equation, one can see that the relation (\ref{9}) can be
written as
\begin{equation}
\tilde\Pi_{R{(2)}}^{(\beta)}(\lambda,p) = 
-\frac{8 e^4}{p^2} m^2 I(T/\lambda)
\int d^{2}q\,
n_{F}(|q^{0}|) 
\left(q^{2}+m^2\right)
\delta^{\prime\prime}(q^2-m^2).
\label{11}
\end{equation}
Performing the $q_0$ integration and neglecting
terms of order $(m/T)$, one finds after some calculation that
Eq. (\ref{11}) becomes
\begin{eqnarray}
\tilde\Pi_{R{(2)}}^{(\beta)}(\lambda,p) & = & 
\frac{8 e^4}{p^2} m^2 I(T/\lambda)
\int_0^\infty \frac{d \omega_{ q}}{ \omega_{ q}}
n_{F}(\omega_{ q})\left[\frac{2}{\omega_{ q}^2}-
\frac{3 m^2}{\omega_{ q}^4}\right] + \cdots 
\nonumber \\
& \simeq &
\frac{8 e^4}{p^2} m^2 I(T/\lambda)
\left[\frac{1}{m^2}\left(1-3\times\frac 1 3\right)\right]+\cdots=
0 + \frac{e^4}{p^2} I(T/\lambda)\,
{\cal O}(m/T).
\label{12}
\end{eqnarray}
It is interesting to note how the individual terms, which would
give infrared divergent contributions independent of $m$,
cancel in the complete amplitude.
A similar behavior also occurs when the photon propagator
is uncut. 

The above results show that, in the limit $m\rightarrow 0$, the
infrared mass singularities in the retarded photon
self-energy cancel at finite temperature, at two loops.
Due to the presence of infrared divergences associated with
a massless photon, one cannot directly conclude that 
contributions like those on the right-hand side of 
(\ref{12}) necessarily vanish as $m\rightarrow 0$.
On the other hand, if we resum the theory, we know that
the photon becomes massive so that we can replace the
infrared regulator $\lambda$ by the photon mass in
the theory, namely $m_\gamma=e/\sqrt{\pi}$. 
We then see that all such contributions would indeed
cancel in the limit $m\rightarrow 0$, which is consistent
with our earlier analysis \cite{brandt:1998gb}.
In view of this behavior of the infrared domain and
due to the presence of an overall factor of $m^2$
in all other regions,
the complete thermal amplitudes in QED$_2$ will therefore
reduce, as $m\rightarrow 0$, to those
of the massless Schwinger model. Presumably, this
occurs because in the transition from a massive to
a massless fermion, the number of fermionic degrees
of freedom is conserved.
Therefore, as far as the physical (advanced/retarded) amplitudes are
concerned, it appears that in the limit of vanishing fermion mass,
the thermal QED$_2$ theory may describe a free gas of massive bosons 
at finite temperature.

\section{Conclusion}\label{sec5}

In this paper, we have extended our earlier one loop results on the calculation
of temperature dependent parts of retarded Greens functions to two loops. 
We have analyzed the temperature dependent part of the retarded
two and three point functions in scalar field theory and in QED.
We have given a simple representation of these functions
in terms of forward scattering amplitudes of on-shell thermal 
particles. This is calculationally much easier than brute
force calculations and leads to the same result 
as calculated from the imaginary time formalism, thereby establishing 
an explicit
agreement between the two formalisms up to this order. 
From these analyses, we find that the temperature dependent part of
the retarded Greens functions, at two loops, 
can have a simple description in terms of
two classes of diagrams. The first one consists of simple forward 
scattering amplitudes of two on-shell thermal particles, corresponding to 
terms quadratic in the statistical factor. The second class,
characterized by terms which are linear in the statistical factor,
involves forward scattering amplitudes of a single on-shell thermal
particle.
These contain all possible one loop self-energy and
vertex corrections, which are retarded and have to 
be calculated at zero temperature. As an application of these results,
we have shown how the infrared singularities in the temperature dependent part
of the retarded photon self-energy cancel at two loops in
QED$_{2}$, in the limit of vanishing fermion mass.

We expect that the above representation of retarded Greens functions
at finite temperature can be generalized for any theory
to $N$-loop order. Namely, these functions may be described in terms
of forward scattering amplitudes of $N$ on-shell thermal
particles, amplitudes of $N-1$ on-shell thermal particles with all
possible retarded one loop self-energy and vertex corrections evaluated
at zero temperature, and so on. Such a representation leads to 
a simple and systematic way of calculating retarded (advanced)
thermal amplitudes, which is rather useful especially in the case
of non-Abelian gauge theories \cite{frenkel:1991tsbrandt:1998}.

\acknowledgements   

A.D. is supported  in  part    by US  DOE  Grant number  DE-FG-02-91ER40685
and NSF-INT-9602559.  F.T.B and J.F. are partially
supported by CNPq (the National Research Council of Brazil) and 
Fapesp. F.T.B is supported in part by PRONEX.
\vfill


\begin{thebibliography}{10}

\bibitem{gross:1981br}
D.~J. Gross, R.~D. Pisarski, and L.~G. Yaffe, Rev. Mod. Phys. {\bf 53},  43
  (1981).

\bibitem{weldon:1982aqweldon:1983jn}
H.~A. Weldon, Phys. Rev. {\bf D26},  1394  (1982);
{\bf D28},  2007  (1983).

\bibitem{kajantie:1985xx}
K. Kajantie and J. Kapusta, Ann. Phys. {\bf 160},  477  (1985).

\bibitem{braaten:1990mzbraaten:1990az}
E. Braaten and R.~D. Pisarski, Nucl. Phys. {\bf B337},  569  (1990)
; {\bf B339},  310  (1990);\\ Phys. Rev.
{\bf D45}, 1827 (1992).

\bibitem{smilga:1996cm}
A.~V. Smilga, Phys. Rept. {\bf 291},  1  (1997).

\bibitem{kapusta:book89}
J.~I. Kapusta, {\em Finite Temperature Field Theory} (Cambridge University
  Press, Cambridge, England, 1989).

\bibitem{lebellac:book96}
M.~L. Bellac, {\em Thermal Field Theory} (Cambridge University Press,
  Cambridge, England, 1996).

\bibitem{das:book97}
A. Das, {\em Finite Temperature Field Theory} (World Scientific, NY, 1997).

\bibitem{landsman:1987uw}
N.~P. Landsman and C.~G. van Weert, Phys. Rept. {\bf 145},  141  (1987).

\bibitem{evans:1992kyevans:1993ak}
T.~S. Evans, Nucl. Phys. {\bf B374},  340  (1992);
Phys. Rev. {\bf D47},  4196  (1993).

\bibitem{baier:1994yh}
R. Baier and A. Niegawa, Phys. Rev. {\bf D49},  4107  (1994).

\bibitem{carrington:1996rx}
M.~E. Carrington and U. Heinz, Eur. Phys. J. {\bf C1},  619  (1998).

\bibitem{brandt:1998gb}
F.~T. Brandt, A. Das, J. Frenkel, and A.~J. da~Silva,   
hep-th/9809177, to be published in Phys. Rev. D.

\bibitem{schwinger:1962tp}
J. Schwinger, Phys. Rev. {\bf 128},  2425  (1962).

\bibitem{parwani:1992gq}
R.~R. Parwani, Phys. Rev. {\bf D45},  4695  (1992).

\bibitem{frenkel:1991tsbrandt:1998}
J. Frenkel and J.~C. Taylor, Nucl. Phys. {\bf B374},  156  (1992);
F.~T. Brandt and J. Frenkel, Phys. Rev. {\bf D58},  085012  (1998).

\end{thebibliography}

\end{document}